\documentclass[12pt]{article}
\usepackage{amssymb}
\usepackage{latexsym}

\usepackage{amsmath}

\input{tcilatex}

\begin{document}

\title{The Form Factors and Quantum Equation of Motion in the sine-Gordon
Model\thanks{
Talk given by H. Babujian at NATO Advanced Research Workshop ``Dynamical
Symmetries in Integrable Two-Dimensional Quantum Field Theories and Lattice
Models'' 25-30 September, Kiev Ukraine.} }
\author{H. Babujian\thanks{%
Permanent address: Yerevan Physics Institute, Alikhanian Brothers 2,
Yerevan, 375036 Armenia.} \thanks{
e-mail: babujian@lx2.yerphi.am, babujian@physik.fu-berlin.de} and M. Karowski%
\thanks{
e-mail: karowski@physik.fu-berlin.de} \\
Institut f\"ur Theoretische Physik\\
Freie Universit\"at Berlin,\\
Arnimallee 14, 14195 Berlin, Germany\\
}
\date{}
\maketitle

\begin{abstract}
Using the methods of the ``form factor program'' exact expressions of all
matrix elements are obtained for several operators of the quantum
sine-Gordon model alias the massive Thirring model. A general formula is
presented which provides form factors in terms of an integral
representation. In particular charge-less operators as for example the
current of the topological charge, the energy momentum tensor and all higher
currents are considered. In the breather sector it is found the quantum
sine-Gordon field equation holds with an exact relation between the ``bare''
mass and the normalized mass. Also a relation for the trace of the energy
momentum is obtained. All results are compared with Feynman graph expansion
and full agreement is found. \\[8pt]
PACS: 11.10.-z; 11.10.Kk; 11.55.Ds\newline
Keywords: Integrable quantum field theory, Form factors
\end{abstract}

\section{Introduction}

This work continues a previous investigation \cite{BFKZ} on exact form
factors for the sine-Gordon alias the massive Thirring model. Coleman \cite
{Co} had shown that these two models are equivalent on the quantum level.
The corresponding classical models are defined by their Lagrangian's 
\begin{eqnarray*}
\mathcal{L}^{SG} &=&\tfrac{1}{2}\partial _{\mu }\varphi \partial ^{\mu
}\varphi +\frac{\alpha }{\beta ^{2}}\left( \cos \beta \varphi -1\right)  \\
\mathcal{L}^{MT} &=&\overline{\psi }(i\gamma \partial -M)\psi -\tfrac{1}{2}%
g\,j^{\mu }j_{\mu }\,\,,\quad \left( j^{\mu }=\overline{\psi }\gamma ^{\mu
}\psi \right) .
\end{eqnarray*}
We do not use these classical Lagrangians and any quantization procedure to
construct the quantum models. We have contact with the classical models
only, when at the end we compare our exact results with Feynman graph
expansions which are based on the Lagrangians. The `form factor program' is
part of the `Bootstrap program for integrable quantum field theories in
1+1-dimensions'. This program classifies integrable quantum field theoretic
models and in addition it provides their explicit exact solutions in term of
all Wightman functions. These results are obtained in three steps:

\begin{enumerate}
\item  The S-matrix is calculated by means of general properties such as
unitarity and crossing, the Yang-Baxter equations (which are a consequence
of integrability) and the additional assumption of `maximal analyticity'.
This means that the two-particle S-matrix is an analytic function in the
physical plane (of the Mandelstam variable $(p_{1}+p_{2})^{2}$) and
possesses only those poles there which are of physical origin.

\item  Generalized form factors which are matrix elements of local operators 
\[
^{out}\left\langle \,p_{m}^{\prime },\ldots ,p_{1}^{\prime }\left| \mathcal{O%
}(x)\right| p_{1},\ldots ,p_{n}\,\right\rangle ^{in}\,
\]
are calculated by means of the S-matrix. More precisely, the equations $%
(i)-(v)$ given below on page \pageref{pf} are used as an input. These
equations follow from LSZ-assumptions and again the additional assumption of
`maximal analyticity' (see also \cite{BFKZ}).

\item  The Wightman functions are obtained by inserting a complete set of
intermediate states. In particular the two point function for a hermitian
operator $\mathcal{O}(x)$ reads 
\[
\langle \mathcal{O}(x)\,\mathcal{O}(0)\rangle =\sum_{n=0}^{\infty }\frac{1}{%
n!}\int \dots \int \frac{dp_{1}\ldots dp_{n}}{\,(2\pi )^{n}2\omega _{1}\dots
2\omega _{n}}\left| \left\langle \,0\left| \mathcal{O}(0)\right|
p_{1},\ldots ,p_{n}\,\right\rangle ^{in}\right| ^{2}e^{-ix\sum p_{i}}. 
\]
\end{enumerate}

The on-shell program i.e. the exact determination of the scattering matrix
was formulated in \cite{KTTW}. Off-shell considerations were carried out in 
\cite{W} and in \cite{KW}, where the concept of a generalized form factor
was introduced and consistency equations were formulated which are expected
to be satisfied by these objects. Thereafter this approach was developed
further and studied in the context of several explicit models (see e.g. \cite
{Sm}\footnote{%
For more references see \cite{BFKZ}.}). More recent papers on solitonic
matrix elements in the sine-Gordon model are \cite{Lu,LZ}. There is a nice
application \cite{GNT,CET} of form factors in condensed matter physics. The
one dimensional Mott insulators can be described in terms of the quantum
sine-Gordon model.

In the previous paper \cite{BFKZ} an integral representation for general
matrix elements of the fundamental fermi-field of the massive Thirring model
has been proposed. In \cite{BK,BK1,BK2} we generalize this formula and
investigate in particular charge-less local operators. The strategy is as
follows:

For a state of $n$ particles of kind $\alpha _{i}$ with momenta $%
p_{i}=m\sinh \theta _{i}$ and a local operator $\mathcal{O}(x)$ the
generalized form factor is defined by 
\begin{equation}
\langle \,0\,|\,\mathcal{O}(x)\,|\,\alpha _{1}(p_{1}),\dots ,\alpha
_{n}(p_{n})\,\rangle ^{in}=e^{-ix(p_{1}+\cdots +p_{n})}\mathcal{O}_{%
\underline{\alpha }}(\underline{\theta })  \label{f}
\end{equation}
for $\theta _{1}>\dots >\theta _{n}$. The short notation $\underline{\alpha }%
=(\alpha _{1},\dots ,\alpha _{n})$ and $\underline{\theta }=(\theta
_{1},\dots ,\theta _{n})$ has been used. We make the Ansatz 
\[
\mathcal{O}_{\underline{\alpha }}(\underline{\theta })=\int_{\mathcal{C}_{%
\underline{\theta }}}dz_{1}\cdots \int_{\mathcal{C}_{\underline{\theta }%
}}dz_{m}\,h(\underline{\theta },{\underline{z}})\,p^{\mathcal{O}}(\underline{%
\theta },{\underline{z}})\,\Psi _{\underline{\alpha }}(\underline{\theta },{%
\underline{z}}) 
\]
with the Bethe state $\Psi _{\underline{\alpha }}(\underline{\theta },{%
\underline{z}})$ defined by eq.~(\ref{2.4}) and the integration contours $%
\mathcal{C}_{\underline{\theta }}$ of figure~\ref{f5.1}. The scalar function 
$h(\underline{\theta },{\underline{z}})$ is uniquely determined by the
S-matrix whereas the scalar `p-function' $p^{\mathcal{O}}(\underline{\theta }%
,{\underline{z})}$ depends on the operator. By means of the Ansatz we
transform the properties $(i)-(v)$ of the co-vector valued function $%
\mathcal{O}_{\underline{\alpha }}(\underline{\theta })$ (see page \pageref
{pf}) to properties $(i^{\prime })-(v^{\prime })$ of the scalar function $p^{%
\mathcal{O}}(\underline{\theta },{\underline{z})}$ which are easily solved.
For example we obtain the p-functions for the local operator\footnote{%
The symbol $\mathcal{N}$ refers to normal products of local quantum fields.} 
$\mathcal{N}\left[ \overline{\psi }\psi \right] (x)$ as 
\[
p^{\overline{\psi }\psi }(\underline{\theta },\underline{z})=N_{n}^{%
\overline{\psi }\psi }\left( \sum\limits_{i=1}^{n}e^{-\theta
_{i}}\sum\limits_{i=1}^{m}e^{z_{i}}-\sum\limits_{i=1}^{n}e^{\theta
_{i}}\sum\limits_{i=1}^{m}e^{-z_{i}}\right) \,. 
\]
In section \ref{s41} we propose in addition the p-functions for $\mathcal{N}%
\left[ \overline{\psi }\gamma ^{5}\psi \right] (x)$, the current $j^{\mu
}(x) $, the energy momentum tensor $T^{\mu \nu }(x)$ and the infinitely many
higher conserved currents $J_{L}^{\mu }(x)$. The identification with the
operators is made by comparing the exact results with Feynman graph
expansions. Properties as charge, behavior under Lorentz transformations
etc. will also become obvious.

\section{Recall of formulae}

\label{s2}

In this section we recall some formulae which we shall need in the following
sections to present our results. All this material can be found in \cite
{BFKZ} including the original references.

\subsection{The S-matrix}

The sine-Gordon model alias massive Thirring model describes the interaction
of several types of particles: solitons, anti-solitons alias fermions and
anti-fermions and a finite number of charge-less breathers, which may be
considered as bound states of solitons and anti-solitons. Integrability of
the model implies that the n-particle S-matrix factorizes into two particle
S-matrices. In particular scattering conserves the number of particles and
even their momenta. The two particle soliton-soliton amplitude $a(\theta )$,
the soliton anti-solitons forward and backward amplitudes $b(\theta )$ and $%
c(\theta )$ are given by \cite{Z}\cite{KTTW} 
\begin{gather}
b(\theta )=\frac{\sinh \theta /\nu }{\sinh (i\pi -\theta )/\nu }\,a(\theta
)\,,~~~~c(\theta )=\frac{\sinh i\pi /\nu }{\sinh (i\pi -\theta )/\nu }%
\,a(\theta )\,,  \nonumber \\
a(\theta )=\exp \int_{0}^{\infty }\frac{dt}{t}\,\frac{\sinh \frac{1}{2}%
(1-\nu )t}{\sinh \frac{1}{2}\nu t\,\cosh \frac{1}{2}t}\,\sinh t\frac{\theta 
}{i\pi }\,.  \label{s}
\end{gather}
The parameter $\theta $ is the absolute value of the rapidity difference $%
\theta =|\theta _{1}-\theta _{2}|$ where $\theta _{i}$ are the rapidities of
the particles given by the momenta $p_{i}=M\sinh \theta _{i}$. The parameter 
$\nu $ is related to the sine-Gordon and massive Thirring model coupling
constant by 
\[
\nu =\frac{\beta ^{2}}{8\pi -\beta ^{2}}=\frac{\pi }{\pi +2g} 
\]
where the second equality is due to Coleman \cite{Co}.

We list some general properties of the two-particle S-matrix. As usual in
this context we use in the notation 
\[
v^{1\dots n}\in V^{1\dots n}=V_{1}\otimes \cdots \otimes V_{n} 
\]
for a vector in a tensor product space. The vector components are denoted by 
$v^{\alpha _{1}\dots \alpha _{n}}$. A linear operator connecting two such
spaces with matrix elements $A_{\alpha _{1}\dots \alpha _{n}}^{\alpha
_{1}^{\prime }\dots \alpha _{n^{\prime }}^{\prime }}$ is denoted by 
\[
A_{1\dots n}^{1^{\prime }\dots n^{\prime }}:~V^{1\dots n}\rightarrow
V^{1^{\prime }\dots n^{\prime }} 
\]
where we omit the upper indices if they are obvious. All vector spaces $%
V_{i} $ are isomorphic to a space $V$ whose basis vectors are label all
kinds of particles (here solitons and anti-solitons, i.e. $V\cong \Bbb{C}%
^{2} $). An S-matrix as $S_{ij}$ acts nontrivial only on the factors $%
V_{i}\otimes V_{j}$.

The physical S-matrix in the formulas above is given for positive values of
the rapidity parameter $\theta $. For later convenience we will also
consider an auxiliary matrix $\dot{S}(\theta _{1},\theta _{2})$ regarded as
a function depending on the individual rapidities of both particles $\theta
_{1},\theta _{2}$ or some times also on the difference $\theta _{1}-\theta
_{2}$%
\[
\dot{S}_{12}(\theta _{1},\theta _{2})=\dot{S}_{12}(\theta _{1}-\theta
_{2})=\left\{ 
\begin{array}{lll}
(\sigma S)_{12}(|\theta _{1}-\theta _{2}|) & \text{for} & \theta _{1}>\theta
_{2} \\ 
(S\sigma )_{21}^{-1}(|\theta _{1}-\theta _{2}|) & \text{for} & \theta
_{1}<\theta _{2}
\end{array}
\right. 
\]
with $\sigma $ taking into account the statistics of the particles. It is a
diagonal matrix $\sigma _{12}$ with entries $-1$ if both particles are
fermions and $+1$ otherwise. The matrix $\dot{S}(\theta _{1},\theta _{2})$
is an analytic function in terms of both variables $\theta _{1}$ and $\theta
_{2}$. The auxiliary matrix $\dot{S}_{12}$ acts nontrivial on the factors $%
V_{1}\otimes V_{2}$ and in addition exchanges these factors, e.g. 
\[
\dot{S}_{12}(\theta )\,:\,V_{1}\otimes V_{2}\rightarrow V_{2}\otimes
V_{1}\,. 
\]
It may be depicted as 
\[
\dot{S}_{12}(\theta _{1},\theta _{2})~~=~~ 
\begin{array}{c}
\unitlength3mm\begin{picture}(5,4) \put(1,0){\line(1,1){4}}
\put(5,0){\line(-1,1){4}} \put(0,.6){$\theta_1$} \put(4.8,.6){$\theta_2$}
\end{picture}
\end{array}
\]
Here and in the following we associate a rapidity variable $\theta _{i}\in %
\mathbb{C}$ to each space $V_{i}$ which is graphically represented by a line
labeled by $\theta _{i}$ or simply by $i$. In terms of the auxiliary
S-matrix the Yang-Baxter equation has the general form 
\[
\dot{S}_{12}(\theta _{12})\,\dot{S}_{13}(\theta _{13})\,\dot{S}_{23}(\theta
_{23})=\dot{S}_{23}(\theta _{23})\,\dot{S}_{13}(\theta _{13})\,\dot{S}%
_{12}(\theta _{12}) 
\]
which graphically simply reads 
\[
\begin{array}{c}
\unitlength6mm\begin{picture}(9,4) \put(0,1){\line(1,1){3}}
\put(0,3){\line(1,-1){3}} \put(2,0){\line(0,1){4}} \put(4.3,2){$=$}
\put(6,0){\line(1,1){3}} \put(6,4){\line(1,-1){3}} \put(7,0){\line(0,1){4}}
\put(.2,.5){$1$} \put(1.3,0){$2$} \put(3,.2){$3$} \put(5.5,.2){$1$}
\put(7.3,0){$2$} \put(8.4,.4){$3$} \end{picture}~~~.
\end{array}
\]
Unitarity and crossing may be written and depicted as 
\begin{equation}
\dot{S}_{21}(\theta _{21})\dot{S}_{12}(\theta _{12})=1~:~~~~~ 
\begin{array}{c}
\unitlength3mm\begin{picture}(8,5) \put(0,1){\line(1,1){2}}
\put(2,1){\line(-1,1){2}} \put(0,3){\line(1,1){2}} \put(2,3){\line(-1,1){2}}
\put(6,1){\line(0,1){4}} \put(8,1){\line(0,1){4}} \put(3.7,2.7){$=$}
\put(0,-.5){$1$} \put(1.5,-.5){$2$} \put(6,-.5){$1$} \put(7.5,-.5){$2$}
\end{picture}
\end{array}
\label{1.8}
\end{equation}
\begin{gather}
\dot{S}_{12}(\theta _{1}-\theta _{2})=\mathbf{C}^{2\bar{2}}\,\dot{S}_{\bar{2}%
1}(\theta _{2}+i\pi -\theta _{1})\,\mathbf{C}_{\bar{2}2}=\mathbf{C}^{1\bar{1}%
}\,\dot{S}_{2\bar{1}}(\theta _{2}-(\theta _{1}-i\pi )\,\mathbf{C}^{\bar{1}1}
\label{1.9} \\
\begin{array}{c}
\unitlength3mm\begin{picture}(4,5) \put(0,1){\line(1,1){4}}
\put(4,1){\line(-1,1){4}} \put(0,-.5){$1$} \put(3.7,-.5){$2$} \end{picture}
\end{array}
~~~~~~=~~~~~~ 
\begin{array}{c}
\unitlength3mm\begin{picture}(6,5) \put(1,1){\line(1,1){4}}
\put(4,1){\line(-1,2){2}} \put(1,5){\oval(2,8)[lb]}
\put(5,1){\oval(2,8)[tr]} \put(3.5,-.5){$1$} \put(5.7,-.5){$2$} \end{picture}
\end{array}
~~~~~~=~~~~~~ 
\begin{array}{c}
\unitlength3mm\begin{picture}(6,5) \put(2,1){\line(1,2){2}}
\put(5,1){\line(-1,1){4}} \put(1,1){\oval(2,8)[lt]}
\put(5,5){\oval(2,8)[br]} \put(0,-.5){$1$} \put(2,-.5){$2$} \end{picture}
\end{array}
~~~~~  \nonumber
\end{gather}
where $\mathbf{C}^{1\bar{1}}$ and $\mathbf{C}_{1\bar{1}}$ are charge
conjugation matrices. For the sine-Gordon model the matrix elements are $%
\mathbf{C}^{\alpha \bar{\beta}}=\mathbf{C}_{\alpha \bar{\beta}}=\delta
_{\alpha \beta }$ where $\bar{\beta}$ denotes the anti-particle of $\beta $.
We have introduced the graphical rule, that a line changing the ``time
direction'' also interchanges particles and anti-particles and changes the
rapidity as $\theta \rightarrow \theta \pm i\pi $, as follows 
\[
\mathbf{C}_{\alpha \bar{\beta}}= 
\begin{array}{c}
\unitlength4mm\begin{picture}(7,3) \put(2,1){\oval(2,4)[t]}
\put(0,1){$\theta$} \put(3.3,1){$\theta-i\pi$} \put(.7,0){$\alpha$}
\put(2.7,-.1){$\bar\beta$} \end{picture}
\end{array}
,~~~~\mathbf{C}^{\alpha \bar{\beta}}= 
\begin{array}{c}
\unitlength4mm\begin{picture}(7,3) \put(2,2){\oval(2,4)[b]}
\put(0,1){$\theta$} \put(3.3,1){$\theta+i\pi$} \put(.7,2.2){$\alpha$}
\put(2.7,2.2){$\bar\beta$} \end{picture}
\end{array}
\;. 
\]
Similar crossing relations will be used below to investigate the properties
of form factors.

Finally we denote a property of the two-particle S-matrix 
\[
\dot{S}_{\alpha \beta }^{\delta \gamma }(0)=-\delta _{\alpha }^{\delta
}\delta _{\beta }^{\gamma } 
\]
which turns out to be true for all examples. This means that $\dot{S}$ for
zero momentum difference is equal to minus the permutation operator.

\subsection{Form factors}

For a state of $n$ particles of kind $\alpha _{i}$ with momenta $p_{i}$ and
a local operator $\mathcal{O}(x)$ we define the form factor functions $%
\mathcal{O}_{\alpha _{1},\dots ,\alpha _{n}}(\theta _{1},\dots ,\theta _{n})$
by eq. (\ref{f}) on page \pageref{f} for the specific order of the rapidities%
$\;\theta _{1}>\dots >\theta _{n}$. For all other arrangements of the
rapidities the functions $\mathcal{O}_{\underline{\alpha }}({\underline{%
\theta }})$ are given by analytic continuation. Note that in general this
analytic continuation does \underline{not} provide the physical values of
the form factor. These are given for ordered rapidities as indicated above
and for other orders by the statistics of the particles, of course. The $%
\mathcal{O}_{\underline{\alpha }}({\underline{\theta }})$ are considered as
the components of a co-vector valued function $\mathcal{O}_{1\dots n}({%
\underline{\theta }})\in V_{1\dots n}=\left( V^{1\dots n}\right) ^{\dagger }$%
which may be depicted as 
\[
\mathcal{O}_{1\dots n}({\underline{\theta }})= 
\begin{array}{c}
\unitlength4mm\begin{picture}(6,4) \put(3,2){\oval(6,2)}
\put(3,2){\makebox(0,0){${\cal O}$}} \put(1,0){\line(0,1){1}}
\put(5,0){\line(0,1){1}} \put(0,0){$\theta_1$} \put(5.3,0){$\theta_n$}
\put(2.5,.5){$\dots$} \end{picture}
\end{array}
. 
\]

Now we formulate the main properties of form factors in terms of the
functions $\mathcal{O}_{1\dots n}({\underline{\theta }})$ which follow from
general LSZ-assumptions and ``maximal analyticity''. The later condition
means that $\mathcal{O}_{1\dots n}({\underline{\theta }})$ is a meromorphic
function with respect to all $\theta $'s and all poles in the `physical'
strips $0<\limfunc{Im}\theta _{ij}<\pi $~$(\theta _{ij}=\theta _{i}-\theta
_{j}\,i<j)$ are of physical origin, as for example bound state poles.

\paragraph{\textbf{Properties:\label{pf}}}

The co-vector valued function $\mathcal{O}_{1\dots n}({\underline{\theta }})$
is meromorphic with respect to all variables $\theta _{1},\dots ,\theta _{n}$
and

\begin{itemize}
\item[$(i)$]  it satisfies the symmetry property under the permutation of
both, the variables $\theta _{i},\theta _{j}$ and the spaces $i,j$ at the
same time 
\[
\mathcal{O}_{\dots ij\dots }(\dots ,\theta _{i},\theta _{j},\dots )=\mathcal{%
O}_{\dots ji\dots }(\dots ,\theta _{j},\theta _{i},\dots )\,\dot{S}%
_{ij}(\theta _{i}-\theta _{j}) 
\]
for all possible arrangements of the $\theta $'s,

\item[$(ii)$]  it satisfies the periodicity property under the cyclic
permutation of the rapidity variables and spaces 
\[
\mathcal{O}_{1\dots n}(\theta _{1},\theta _{2},\dots ,\theta _{n},)=\mathcal{%
O}_{2\dots n1}(\theta _{2},\dots ,\theta _{n},\theta _{1}-2\pi i)\sigma _{%
\mathcal{O}1} 
\]

\item[$(iii)$]  and it has poles determined by one-particle states in each
sub-channel. In particular the function $\mathcal{O}_{\underline{\alpha }}({%
\underline{\theta }})$ has a pole at $\theta _{12}=i\pi $ such that 
\[
\limfunc{Res}\limits_{\theta _{12}=i\pi }\mathcal{O}_{1\dots n}(\theta
_{1},\dots ,\theta _{n})=2i\,\mathbf{C}_{12}\,\mathcal{O}_{3\dots n}(\theta
_{3},\dots ,\theta _{n})\left( \mathbf{1}-S_{2n}\dots S_{23}\right) 
\]
where $\mathbf{C}_{12}$ is the charge conjugation matrix.

\item[$(iv)$]  \label{iv}If the model also possesses bound states, the
function $\mathcal{O}_{\underline{\alpha }}({\underline{\theta }})$ has
additional poles. If for instance the particles 1 and 2 form a bound state
(12), there is a pole at $\theta _{12}=iu_{12}^{(12)}~(0<u_{12}^{(12)}<\pi )$
such that 
\[
\limfunc{Res}_{\theta _{12}=iu_{12}^{(12)}}\mathcal{O}_{12\dots n}(\theta
_{1},\theta _{2},\dots ,\theta _{n})\,=\mathcal{O}_{(12)\dots n}(\theta
_{(12)},\dots ,\theta _{n})\,\sqrt{2}\Gamma _{12}^{(12)}\, 
\]
where the bound state intertwiner $\Gamma _{12}^{(12)}$ and the relations of
the rapidities $\theta _{1},\theta _{2},\theta _{(12)}$ and the fusion angle 
$u_{12}^{(12)}$ \cite{BK}.

\item[$(v)$]  Naturally, since we are dealing with relativistic quantum
field theories we finally have Lorentz covariance 
\[
\mathcal{O}_{1\dots n}(\theta _{1}+\mu ,\dots ,\theta _{n}+\mu )=e^{s\mu }\,%
\mathcal{O}_{1\dots n}(\theta _{1},\dots ,\theta _{n}) 
\]
if the local operator transforms as $\mathcal{O}\to e^{s\mu }\mathcal{O}$
where $s$ is the ``spin'' of $\mathcal{O}$.
\end{itemize}

In the formulae $(i)$ the statistics of the particles is taken into account
by $\dot{S}$ which means that $\dot{S}_{12}=-S_{12}$ if both particles are
fermions and $\dot{S}_{12}=S_{12}$ otherwise. In $(ii)$ the statistics of
the operator $\mathcal{O}$ is taken into account by $\sigma _{\mathcal{O}%
1}=-1$ if both the operator $\mathcal{O}$ and particle 1 are fermionic and $%
\sigma _{\mathcal{O}1}=1$ otherwise.

The property $(i)-(iv)$ may be depicted as 
\[
\begin{array}{rrcl}
(i) & 
\begin{array}{c}
\unitlength4mm\begin{picture}(7,3) \put(3.5,2){\oval(7,2)}
\put(3.5,2){\makebox(0,0){${\cal O}$}} \put(1,0){\line(0,1){1}}
\put(3,0){\line(0,1){1}} \put(4,0){\line(0,1){1}} \put(6,0){\line(0,1){1}}
\put(1.4,.5){$\dots$} \put(4.4,.5){$\dots$} \end{picture}
\end{array}
& = & 
\begin{array}{c}
\unitlength4mm\begin{picture}(7,4) \put(3.5,3){\oval(7,2)}
\put(3.5,3){\makebox(0,0){${\cal O}$}} \put(1,0){\line(0,1){2}}
\put(3,0){\line(1,2){1}} \put(4,0){\line(-1,2){1}} \put(6,0){\line(0,1){2}}
\put(1.4,1){$\dots$} \put(4.4,1){$\dots$} \end{picture}
\end{array}
\\ 
(ii) & 
\begin{array}{c}
\unitlength4mm\begin{picture}(5,4) \put(2.5,2){\oval(5,2)}
\put(2.5,2){\makebox(0,0){${\cal O}$}} \put(1,0){\line(0,1){1}}
\put(2,0){\line(0,1){1}} \put(4,0){\line(0,1){1}} \put(2.4,.5){$\dots$}
\end{picture}
\end{array}
& = & 
\begin{array}{c}
\unitlength4mm\begin{picture}(7,4) \put(0,0){\line(0,1){1}}
\put(6,1){\oval(2,2)[b]} \put(3.5,1){\oval(7,6)[t]} \put(3.5,2){\oval(5,2)}
\put(3.5,2){\makebox(0,0){${\cal O}$}} \put(2,0){\line(0,1){1}}
\put(4,0){\line(0,1){1}} \put(2.4,.5){$\dots$} \end{picture}
\end{array}
\sigma _{\mathcal{O}1} \\ 
(iii) & \frac{1}{2i}\,\limfunc{Res}\limits_{\theta _{12}=i\pi } 
\begin{array}{c}
\unitlength4mm\begin{picture}(6,4) \put(3,2){\oval(6,2)}
\put(3,2){\makebox(0,0){${\cal O}$}} \put(1,0){\line(0,1){1}}
\put(2,0){\line(0,1){1}} \put(3,0){\line(0,1){1}} \put(5,0){\line(0,1){1}}
\put(3.4,.5){$\dots$} \end{picture}
\end{array}
& = & 
\begin{array}{c}
\unitlength4mm\begin{picture}(5,4) \put(.5,0){\oval(1,2)[t]}
\put(3,2){\oval(4,2)} \put(3,2){\makebox(0,0){${\cal O}$}}
\put(2,0){\line(0,1){1}} \put(4,0){\line(0,1){1}} \put(2.4,.5){$\dots$}
\end{picture}
\end{array}
- 
\begin{array}{c}
\unitlength4mm\begin{picture}(6,5) \put(0,0){\line(0,1){3}}
\put(3,3){\oval(6,4)[t]} \put(3,3){\oval(6,4)[br]} \put(3,0){\oval(4,2)[tl]}
\put(3,3){\oval(4,2)} \put(3,3){\makebox(0,0){${\cal O}$}}
\put(2,0){\line(0,1){2}} \put(4,0){\line(0,1){2}} \put(2.4,1.5){$\dots$}
\end{picture}
\end{array}
\sigma _{\mathcal{O}1} \\ 
(iv) & \frac{1}{\sqrt{2}}\limfunc{Res}\limits_{\theta _{12}=iu_{12}^{(12)}} 
\begin{array}{c}
\unitlength4mm\begin{picture}(5,3) \put(2.5,2){\oval(5,2)}
\put(2.5,2){\makebox(0,0){${\cal O}$}} \put(1,0){\line(0,1){1}}
\put(2,0){\line(0,1){1}} \put(4,0){\line(0,1){1}} \put(2.4,.5){$\dots$}
\end{picture}
\end{array}
& = & 
\begin{array}{c}
\unitlength4mm%
\begin{picture}(5,4) \put(2.5,3){\oval(5,2)}
 \put(2.5,3){\makebox(0,0){${\cal O}$}} \put(1.5,0){\oval(1,2)[t]}
\put(1.5,1){\line(0,1){1}} \put(4,0){\line(0,1){2}} \put(2.4,1){$\dots$}
 \end{picture}
\end{array}
\end{array}
\]

We will now provide a constructive and systematic way of how to solve the
properties $(i)-(v)$ for the co-vector valued function $f$ once the
scattering matrix is given. These solutions are candidates of form factors.
To capture the vectorial structure of the form factors we will employ the
techniques of the algebraic Bethe Ansatz which we now explain briefly.

\subsection{The `off-shell' Bethe Ansatz co-vectors}

As usual in the context of algebraic Bethe Ansatz we define the monodromy
matrix as 
\begin{equation}
T_{1\dots n,0}({\underline{\theta }},\theta _{0})=\dot{S}_{10}(\theta
_{1}-\theta _{0})\,\dot{S}_{20}(\theta _{2}-\theta _{0})\cdots \dot{S}%
_{n0}(\theta _{n}-\theta _{0})= 
\begin{array}{c}
\unitlength3mm\begin{picture}(10,4) \put(0,2){\line(1,0){10}}
\put(2,0){\line(0,1){4}} \put(4,0){\line(0,1){4}} \put(8,0){\line(0,1){4}}
\put(1,0){$1$} \put(3,0){$ 2$} \put(7,0){$ n$} \put(9,.8){$ 0$}
\put(5,1){$\dots$} \end{picture}
\end{array}
.  \nonumber
\end{equation}
It is a matrix acting in the tensor product of the ``quantum space'' $%
V^{1\dots n}=V_{1}\otimes \cdots \otimes V_{n}$ and the ``auxiliary space'' $%
V_{0}$ (all $V_{i}\cong \mathbb{C}^{2}$ = soliton-anti-soliton space). The
sub-matrices $A,B,C,D$ with respect to the auxiliary space are defined by 
\[
T_{1\dots n,0}({\underline{\theta }},z)\equiv \left( 
\begin{array}{cc}
A_{1\dots n}({\underline{\theta }},z) & B_{1\dots n}({\underline{\theta }},z)
\\ 
C_{1\dots n}({\underline{\theta }},z) & D_{1\dots n}({\underline{\theta }},z)
\end{array}
\right) \,. 
\]
A Bethe Ansatz co-vector in $V_{1\dots n}$ is given by 
\begin{equation}
\begin{array}{rcl}
\Psi _{1\dots n}({\underline{\theta }},\underline{z}) & = & \Omega _{1\dots
n}C_{1\dots n}({\underline{\theta }},z_{1})\cdots C_{1\dots n}({\underline{%
\theta }},z_{m}) \\ 
\begin{array}{c}
\unitlength5mm\begin{picture}(6,4) \put(3,2){\oval(6,2)}
\put(3,2){\makebox(0,0){$\Psi$}} \put(1,0){\line(0,1){1}}
\put(5,0){\line(0,1){1}} \put(0,0){$\theta_1$} \put(5.3,0){$\theta_n$}
\put(2.5,.5){$\dots$} \end{picture}
\end{array}
& = & 
\begin{array}{c}
\unitlength5mm\begin{picture}(6,4.5) \put(0,1){\line(1,0){6}}
\put(0,1){\vector(1,0){.5}} \put(6,1){\vector(-1,0){.5}}
\put(0,3){\line(1,0){6}} \put(0,3){\vector(1,0){.5}}
\put(6,3){\vector(-1,0){.5}} \put(1,0){\vector(0,1){4}}
\put(5,0){\vector(0,1){4}} \put(0,0){$\theta_1$} \put(4,0){$\theta_n$}
\put(5.4,.3){$z_m$} \put(5.4,3.3){$z_1$} \put(2.5,2){$\dots$}
\put(.3,1.7){$\vdots$} \put(5.3,1.7){$\vdots$} \end{picture}
\end{array}
\end{array}
\label{2.4}
\end{equation}
where $\underline{z}=(z_{1},\dots ,z_{m})$. Usually one has the restriction $%
2m\leq n$ and the charge of the state is $q=n-2m=$ number of solitons minus
number of anti-solitons. The solitons are depicted by $\uparrow $ or $%
\leftarrow $ and anti-solitons by $\downarrow $ or $\rightarrow $. The
co-vector $\Omega _{1\dots n}$ is the ``pseudo-vacuum'' consisting only of
solitons (highest weight states) 
\[
\Omega _{1\dots n}=\uparrow \otimes \cdots \otimes \uparrow \,. 
\]
It satisfies 
\[
\begin{array}{rcl}
\Omega _{1\dots n}\,B_{1\dots n}({\underline{\theta }},z) & = & 0 \\ 
\Omega _{1\dots n}\,A_{1\dots n}({\underline{\theta }},z) & = & 
\prod\limits_{i=1}^{n}\dot{a}(\theta _{i}-z)\Omega _{1\dots n} \\ 
\Omega _{1\dots n}\,D_{1\dots n}({\underline{\theta }},z) & = & 
\prod\limits_{i=1}^{n}\dot{b}(\theta _{i}-z)\Omega _{1\dots n}\,.
\end{array}
\]
The eigenvalues of the matrices $A$ and $D$, i.e. the functions $\dot{a}=-a$
and $\dot{b}=-b$ are given by the amplitudes of the scattering matrix (\ref
{s}). In the following we use the co-vector $\Psi _{1\dots n}({\underline{%
\theta }},\underline{z})$ in its `off-shell' version which means that we do
not fix the parameters $\underline{z}$ by means of Bethe Ansatz equations
but we integrate over the $z$'s.

\section{The general form factor formula}

\label{s3}

In this section we present our main result. We derive a general formula in
terms of an integral representation which allows to construct form factors
i.e. matrix elements of local fields as given by eq.~(\ref{f}). More
precisely, we construct co-vector valued functions which satisfy the
properties $(i)-(v)$ on page \pageref{pf}.

\begin{figure}[tbh]
\[
\unitlength4mm%
\begin{picture}(27,13) \thicklines \put(1,0){
\put(0,0){$\bullet~\theta_n-2\pi i$}
\put(0,2){$\bullet$}\put(.5,1.6){$\theta_n-i\pi\nu$}
\put(.19,3.2){\circle{.3}~$\theta_n-i\pi$} \put(0,6){$\bullet~~\theta_n$}
\put(.2,6.2){\oval(1,1)}\put(-.1,5.71){\vector(-1,0){0}}
\put(.19,7.2){\circle{.3}~$\theta_n+i\pi(\nu-1)$}
\put(0,9){$\bullet~\theta_n+i\pi$}
\put(.19,11.2){\circle{.3}~$\theta_n+i\pi(2\nu-1)$} } \put(8,6){\dots}
\put(12,0){ \put(0,0){$\bullet~\theta_2-2\pi i$}
\put(0,2){$\bullet$}\put(.5,1.6){$\theta_2-i\pi\nu$}
\put(.19,3.2){\circle{.3}~$\theta_2-i\pi$} \put(0,6){$\bullet~~\theta_2$}
\put(.2,6.2){\oval(1,1)}\put(-.1,5.71){\vector(-1,0){0}}
\put(.19,7.2){\circle{.3}~$\theta_2+i\pi(\nu-1)$}
\put(0,9){$\bullet~\theta_2+i\pi$}
\put(.19,11.2){\circle{.3}~$\theta_2+i\pi(2\nu-1)$} } \put(20,1){
\put(0,0){$\bullet~\theta_1-2\pi i$}
\put(0,2){$\bullet$}\put(.5,1.6){$\theta_1-i\pi\nu$}
\put(.19,3.2){\circle{.3}~$\theta_1-i\pi$} \put(0,6){$\bullet~~\theta_1$}
\put(.2,6.2){\oval(1,1)}\put(-.1,5.71){\vector(-1,0){0}}
\put(.19,7.2){\circle{.3}~$\theta_1+i\pi(\nu-1)$}
\put(0,9){$\bullet~\theta_1+i\pi$}
\put(.19,11.2){\circle{.3}~$\theta_1+i\pi(2\nu-1)$} }
\put(9,2.7){\vector(1,0){0}} \put(0,3.2){\oval(34,1)[br]}
\put(27,3.2){\oval(20,1)[tl]} \end{picture}
\]
\caption{\textit{The integration contour $\mathcal{C_{\protect\underline{%
\theta }}}$ (for the repulsive case $\nu >1$). The bullets belong to poles
of the integrand resulting from $u(\theta _{i}-u_{j})\,\phi (\theta
_{i}-u_{j})$ and the small open circles belong to poles originating from $%
t(\theta _{i}-u_{j})$ and $r(\theta _{i}-u_{j})$. }}
\label{f5.1}
\end{figure}
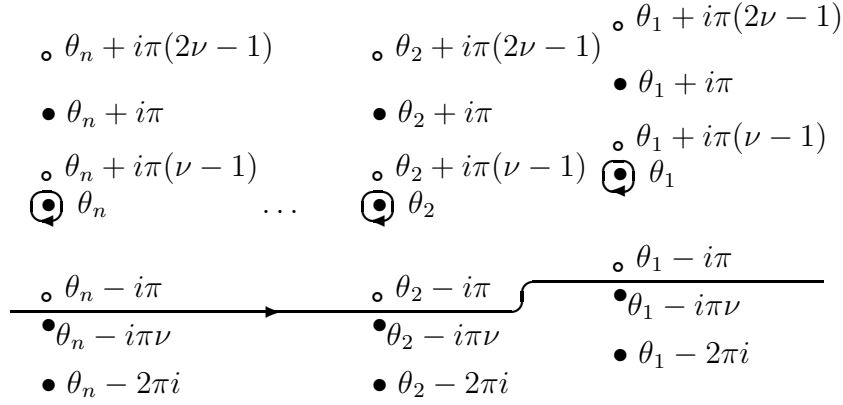

As a candidate of a generalized form factor of a local operator $\mathcal{O}%
(0)$ we make the following Ansatz for the co-vector valued function 
\begin{equation}
\mathcal{O}_{1\dots n}(\underline{\theta })=\int_{\mathcal{C}_{\underline{%
\theta }}}dz_{1}\cdots \int_{\mathcal{C}_{\underline{\theta }}}dz_{m}\,h(%
\underline{\theta },{\underline{z}})\,p^{\mathcal{O}}(\underline{\theta },{%
\underline{z}})\,\Psi _{1\dots n}(\underline{\theta },{\underline{z}})
\label{1.2}
\end{equation}
with the Bethe Ansatz state $\Psi _{1\dots n}(\underline{\theta },{%
\underline{z}})$ defined by eq.~(\ref{2.4}). For all integration variables $%
z_{j}$ $(j=1,\dots ,m)$ the integration contours $\mathcal{C}_{\underline{%
\theta }}$ consists of several pieces (see figure~\ref{f5.1}):

\begin{itemize}
\item[a)]  A line from $-\infty $ to $\infty $ avoiding all poles such that $%
\limfunc{Im}\theta _{i}-\pi -\epsilon <\limfunc{Im}z_{j}<\limfunc{Im}\theta
_{i}-\pi $.

\item[b)]  Clock wise oriented circles around the poles (of the $\phi
(\theta _{i}-z_{j})$) at $z_{j}=\theta _{i}$ $(i=1,\dots ,n)$.
\end{itemize}

Let the scalar function (c.f. \cite{BFKZ}) 
\begin{equation}
h(\underline{\theta },{\underline{z}})=\prod_{1\le i<j\le n}F(\theta
_{ij})\prod_{i=1}^{n}\prod_{j=1}^{m}\phi (\theta _{i}-z_{j})\prod_{1\le
i<j\le m}\tau (z_{i}-z_{j})\,,  \label{1.3}
\end{equation}
be given by 
\begin{equation}
\tau (z)=\frac{1}{\phi (z)\,\phi (-z)}~,~~~~\phi (z)=\frac{1}{F(z)\,F(z+i\pi
)}  \label{1.4}
\end{equation}
and 
\begin{eqnarray}
F(\theta ) &=&\sin \frac{1}{2i}\theta \,\,f_{ss}^{(0)}(\theta )  \nonumber \\
f_{ss}^{(0)}(\theta ) &=&\exp \int_{0}^{\infty }\frac{dt}{t}\frac{\sinh 
\frac{1}{2}(1-\nu )t}{\sinh \frac{1}{2}\nu t\,\cosh \frac{1}{2}t}\frac{%
1-\cosh t(1-\theta /(i\pi ))}{2\sinh t}.  \label{1.7}
\end{eqnarray}
The function $F(\theta )$ is the soliton-soliton form factor fulfilling
Watson's equations 
\begin{equation}
F(\theta )=F(-\theta )\,\dot{a}(\theta )=F(2\pi i-\theta )  \label{2.12}
\end{equation}
with $\,\dot{a}(\theta )=-a(\theta )$ where $a(\theta )$ is the
soliton-soliton scattering amplitude.

\paragraph{Remarks:}

\begin{itemize}
\item  Using Watson's equations (\ref{2.12}) for $F(z)$, crossing (\ref{1.9}%
) and unitarity (\ref{1.8}) for the sine-Gordon amplitudes one derives the
following identities for the scalar functions $\phi (z)$ and $\tau (z)$ from
the definitions (\ref{1.4}) 
\[
\phi (z)=\phi (i\pi -z)=\frac{1}{\dot{b}(z)}\,\phi (z-i\pi )=\frac{\dot{a}%
(z-2\pi i)}{\dot{b}(z)}\,\phi (z-2\pi i)~, 
\]
\[
\tau (z)=\tau (-z)=\frac{b(z)}{a(z)}\,\frac{a(2\pi i-z)}{b(2\pi i-z)}\,\tau
(z-2\pi i) 
\]
where $b(z)$ is the soliton-anti-soliton scattering amplitude related to $%
a(z)$ by crossing $b(z)=a(i\pi -z)$.

\item  The functions $\phi (z)$ and $\tau (z)$ are of the form 
\[
\phi (z)=\frac{const}{\sinh z}\exp \int_{0}^{\infty }\frac{dt}{t}\frac{\sinh 
\frac{1}{2}(1-\nu )t\,\left( \cosh t(\frac{1}{2}-z/(i\pi ))-1\right) }{\sinh 
\frac{1}{2}\,\nu t\sinh t} 
\]
\[
\tau (z)=const.~\sinh z\sinh z/\nu 
\]

\item  The function $h(\underline{\theta },{\underline{z}})$ and the state $%
\Psi _{1\dots n}(\underline{\theta },{\underline{z}})$ are completely
determined by the S-matrix.
\end{itemize}

In contrast to the functions $F(z),\phi (z)$ and $\tau (z)$ the `p-function' 
$p^{\mathcal{O}}(\underline{\theta },{\underline{z}})$ in the integral
representation (\ref{1.2}) depends on the local operator $\mathcal{O}(x),$in
particular on the spin, the charge and the statistics. The number of the
particles $n$ and the number of integrations $m$ are related by $q=n-2m$
where $q$ is the charge of the operator $\mathcal{O}(x)$. The p-functions is
an entire function in the $z_{j}~(j=1,\dots ,m)$ and in order that the form
factor satisfies the properties $(i)-(v)$ it has to satisfy the following

\paragraph{Conditions:\label{p}}

The p-function $p_{n}^{\mathcal{O}}(\underline{\theta },\underline{z})$
(where $n$ is the number of particles and the number of variables $\theta $)
satisfies

\begin{itemize}
\item[$(i^{\prime })$]  $p_{n}^{\mathcal{O}}(\underline{\theta },\underline{z%
})$ is symmetric with respect to the $\theta $'s and the $z$'s.

\item[$(ii^{\prime })$]  $p_{n}^{\mathcal{O}}(\underline{\theta },\underline{%
z})=\sigma _{\mathcal{O}i}p_{n}^{\mathcal{O}}(\dots ,\theta _{i}-2\pi
i,\dots ,\underline{z})$ and it is a polynomial in $e^{\pm z_{j}}~(j=1,\dots
,m)$.

The statistics factor $\sigma _{\mathcal{O}i}$ is $-1$ if the operator $%
\mathcal{O}(x)$ and the particle $i$ are both fermionic and $+1$ otherwise.

\item[$(iii^{\prime })$]  $\left\{ 
\begin{array}{l}
p_{n}^{\mathcal{O}}(\theta _{1}=\theta _{n}+i\pi ,\tilde{\underline{\theta }}%
,\theta _{n};\tilde{\underline{z}},z_{m}=\theta _{n}{)}=\dfrac{\varkappa }{m}%
\,p_{n-2}^{\mathcal{O}}(\tilde{\underline{\theta }},\tilde{\underline{z}})+%
\tilde{p}^{(1)}({\underline{\theta }}) \\ 
p_{n}^{\mathcal{O}}(\theta _{1}=\theta _{n}+i\pi ,\tilde{\underline{\theta }}%
,\theta _{n};\tilde{\underline{z}},z_{m}=\theta _{1}{)}=\sigma _{\mathcal{O}%
1}\dfrac{\varkappa }{m}\,p_{n-2}^{\mathcal{O}}(\tilde{\underline{\theta }},%
\tilde{\underline{z}})+\tilde{p}^{(2)}({\underline{\theta }})
\end{array}
\right. $

where $\tilde{\underline{\theta }}=(\theta _{2},\dots ,\theta _{n-1}),\;%
\tilde{\underline{z}}=(z_{1},\dots z_{m-1})$ and where $\tilde{p}^{(1,2)}({%
\underline{\theta }})$ are independent of the $z$'s and non-vanishing only
for charge less operators $\mathcal{O}(x)$. The constant $\varkappa $
depends on the coupling and is given by (see formula (B.7) in \cite{BFKZ}) 
\[
\varkappa =\frac{\left( f_{ss}^{(0)}(0)\right) ^{2}}{4\pi }\,. 
\]

\item[$(iv^{\prime })$]  the bound state p-functions are investigated in
section \ref{s7}

\item[$(v^{\prime })$]  $p_{n}^{\mathcal{O}}(\underline{\theta }+\mu ,%
\underline{z}+\mu )=e^{s\mu }p_{n}^{\mathcal{O}}(\underline{\theta },%
\underline{z})$ where $s$ is the `spin' of the operator $\mathcal{O}(x)$.
\end{itemize}

As an extension of theorem 4.1 in \cite{BFKZ} we prove the following theorem
which allow to construct generalized form factors.

\begin{theorem}
\label{t} The co-vector valued function $\mathcal{O}_{1\dots n}({\underline{%
\theta }})$ defined by (\ref{1.2}) fulfills the properties $(i),\,(ii)$ and $%
(iii)$ on page \pageref{pf} if the functions $F(\theta ),\allowbreak \phi
(z) $ and $\tau (z)$ are given by definition (\ref{1.3}) -- (\ref{1.7}) and
if the p-function $p_{n}^{\mathcal{O}}(\underline{\theta },\underline{z})$
satisfies the conditions $(i^{\prime })-(iii^{\prime })$.
\end{theorem}

This theorem is proven in \cite{BK}.

\paragraph{Remarks:}

\begin{itemize}
\item  The number of C-operators $m$ depends on the charge $q=n-2m$ of the
operator $\mathcal{O}$, e.g. $m=(n-1)/2$ for the soliton field $\psi (x)$
with charge $q=1$ and $m=n/2$ for charge-less operators like $\overline{\psi 
}\psi $ or the energy momentum tensor $T^{\mu \nu }$.

\item  Note that other sine-Gordon form factors can be calculated from the
general formula (\ref{1.2}) using the bound state formula $(iv)$.

\item  The general representation of form factors by formula (\ref{1.2}) is
not specific to the sine-Gordon model. It may be applied to all integrable
quantum field theoretic model. The main difficulty is to solve the
corresponding Bethe Ansatz.
\end{itemize}

\section{Examples of ``p-functions''}

\label{s41}

In this section we propose the p-functions for various local operators.
Since the charge of the operators which we consider is zero the number of
integrations $m$ and the number of particles $n$ are related by $m=n/2$ and
form factors are non-vanishing only for even number of particles $%
n=2,4,\dots $. We consider p-functions of the form 
\begin{equation}
p_{n}^{\mathcal{O}}(\underline{\theta },{\underline{z}})=N_{n}^{\mathcal{O}%
}\left( p_{+}^{\mathcal{O}}(P^{\mu })\sum_{j=1}^{m}e^{Lz_{j}}+p_{-}^{%
\mathcal{O}}(P^{\mu })\sum_{j=1}^{m}e^{-Lz_{j}}\right)  \label{4.1}
\end{equation}
where $P^{\mu }$ is the total energy momentum vector of all particles. The
integrals in (\ref{1.2}) converge for $L<(1/\nu +1)(n/2-m+1)+1/\nu $. For
large values of $L$ the form factors are defined in general as the analytic
continuations of the integral representation from sufficiently small values
of $\nu $ to other values. Obviously the p-functions (\ref{4.1}) satisfy the
conditions $(i^{\prime })-(iii^{\prime })$ on page \pageref{p}. From the
property $(iii^{\prime })$ we obtain the recursion relation for the
normalization constants 
\[
N_{n}^{\mathcal{O}}=N_{n-2}^{\mathcal{O}}\frac{\varkappa }{m}\quad
\Rightarrow \quad N_{n}^{\mathcal{O}}=N_{2}^{\mathcal{O}}\frac{1}{m!}%
\varkappa ^{m-1} 
\]
where $N_{2}^{\mathcal{O}}$ follows from the two-particle form factors. For
the local operators $\mathcal{N}\left[ \overline{\psi }\psi \right] (x)$, $%
\mathcal{N}\left[ \overline{\psi }\gamma ^{5}\psi \right] (x)$, the current $%
j^{\mu }(x)=\mathcal{N}\left[ \overline{\psi }\gamma ^{\mu }\psi \right] (x)$%
, the energy momentum tensor $T^{\mu \nu }(x)=\tfrac{i}{2}\mathcal{N}\left[ 
\overline{\psi }\gamma ^{\mu }\overleftrightarrow{\partial ^{\nu }\rule%
{0in}{0.14in}}\psi \right] (x)-g^{\mu \nu }\mathcal{L}^{MT}$ and the
infinitely many higher conserved currents $J_{L}^{\mu }(x)$ we propose the
following p-functions 
\begin{eqnarray*}
p^{\overline{\psi }\psi }(\underline{\theta },\underline{z}) &=&N_{n}^{%
\overline{\psi }\psi }\left( \sum\limits_{i=1}^{n}e^{-\theta
_{i}}\sum\limits_{i=1}^{m}e^{z_{i}}-\sum\limits_{i=1}^{n}e^{\theta
_{i}}\sum\limits_{i=1}^{m}e^{-z_{i}}\right) \\
p^{\overline{\psi }\gamma ^{5}\psi }(\underline{\theta },\underline{z})
&=&N_{n}^{\overline{\psi }\gamma ^{5}\psi }\left(
\sum\limits_{i=1}^{n}e^{-\theta
_{i}}\sum\limits_{i=1}^{m}e^{z_{i}}+\sum\limits_{i=1}^{n}e^{\theta
_{i}}\sum\limits_{i=1}^{m}e^{-z_{i}}\right) \\
p^{j^{\pm }}(\underline{\theta },\underline{z}) &=&\pm N_{n}^{j}\left(
\sum\limits_{i=1}^{n}e^{\mp \theta _{i}}\right) ^{-1}\left(
\sum\limits_{i=1}^{n}e^{-\theta
_{i}}\sum\limits_{i=1}^{m}e^{z_{i}}+\sum\limits_{i=1}^{n}e^{\theta
_{i}}\sum\limits_{i=1}^{m}e^{-z_{i}}\right) \\
p^{T^{\pm \pm }}(\underline{\theta },\underline{z})
&=&N_{n}^{T}\sum_{i=1}^{n}e^{\pm \theta _{i}}\left( \sum_{i=1}^{n}e^{\mp
\theta _{i}}\right) ^{-1}\left( \sum\limits_{i=1}^{n}e^{-\theta
_{i}}\sum\limits_{i=1}^{m}e^{z_{i}}-\sum\limits_{i=1}^{n}e^{\theta
_{i}}\sum\limits_{i=1}^{m}e^{-z_{i}}\right) \\
p^{T^{+-}}(\underline{\theta },\underline{z}) &=&-N_{n}^{T}\left(
\sum\limits_{i=1}^{n}e^{-\theta
_{i}}\sum\limits_{i=1}^{m}e^{z_{i}}-\sum\limits_{i=1}^{n}e^{\theta
_{i}}\sum\limits_{i=1}^{m}e^{-z_{i}}\right) \\
p^{J_{L}^{\pm }}(\underline{\theta },\underline{z}) &=&\pm
N_{n}^{J_{L}}\sum\limits_{i=1}^{n}e^{\pm \theta
_{i}}\sum\limits_{i=1}^{m}e^{Lz_{i}}\,,\quad (L=\pm 1,\pm 3,\dots )\,.
\end{eqnarray*}
The identification with the operators has been made by comparing the exact
results with Feynman graph expansions. Properties as charge, behavior under
Lorentz transformations etc. also become obvious \cite{BK} The fundamental
sine-Gordon bose field $\varphi (x)$ which correspond to the lowest breather
is related to the current by Coleman's formula \cite{Co} 
\[
\epsilon ^{\mu \nu }\partial _{\nu }\varphi =-\frac{2\pi }{\beta }j^{\mu
}\quad \text{or}\quad \partial ^{\pm }\varphi =\pm \frac{2\pi }{\beta }%
j^{\pm }. 
\]
This implies for the p-function 
\[
p_{n}^{\varphi }(\underline{\theta },\underline{z})=N_{n}^{j}\frac{2\pi i}{%
\beta M}\left( \sum_{i=1}^{n}e^{\theta }\sum_{i=1}^{n}e^{-\theta }\right)
^{-1}\left( \sum_{i=1}^{n}e^{-\theta
}\sum_{i=1}^{m}e^{z}+\sum_{i=1}^{n}e^{\theta }\sum_{i=1}^{m}e^{-z}\right) . 
\]

\section{Soliton Breather form factors}

\label{s7}

We calculate breather form factors starting with the general formula (\ref
{1.2}) for the soliton form factors using the property $(vi)$ on page 
\pageref{iv} (for details see \cite{BK,BK2}). The $b_{k}$-breather-$(n-2)$%
-soliton form factor is obtained from $\mathcal{O}_{123\dots n}(\underline{%
\theta })$ by means of the fusion procedure $(iv)$ with the fusion angle
given by $u_{12}^{(12)}=u^{(k)}=\pi (1-k\nu )$ the bound state rapidity $\xi
=\theta _{(12)}=\frac{1}{2}(\theta _{1}+\theta _{2})$ and $\underline{\theta 
}^{\prime }=\theta _{3},\dots ,\theta _{n}$ 
\[
\limfunc{Res}_{\theta _{12}=iu^{(k)}}\mathcal{O}_{123\dots n}(\underline{%
\theta })=\,\mathcal{O}_{(12)3\dots n}(\xi ,\underline{\theta }^{\prime })%
\sqrt{2}\Gamma _{12}^{(12)}(iu^{(k)}) 
\]
where the bound state intertwiner \cite{BK}.

For $\theta _{12}\rightarrow iu^{(k)}$ there will be pinchings of the
integration contours in formula (\ref{1.2}) at the poles $%
z_{i}=z^{(l)}=\theta _{2}-i\pi l\nu =\xi -\frac{1}{2}i\pi (1-k\nu +2l\nu )$
for $l=0,\dots ,k$ and $i=1,\dots ,m$. Using the pinching rule of contour
integrals and the symmetry with respect to the $m~z$-integrations we obtain 
\begin{multline*}
\limfunc{Res}_{\theta _{12}=iu^{(k)}}\mathcal{O}_{123\dots n}(\underline{%
\theta })\,=\limfunc{Res}_{\theta _{12}=iu^{(k)}}(-2\pi i)\,m\sum_{l=0}^{k}%
\limfunc{Res}_{z_{1}=z^{(l)}}\int_{\mathcal{C}_{\underline{\theta }%
}}dz_{2}\cdots \int_{\mathcal{C}_{\underline{\theta }}}dz_{m} \\
\times h(\underline{\theta }{,\underline{z}})p^{\mathcal{O}}(\underline{%
\theta }{,\underline{z}})\,\Psi _{1\dots n}(\underline{\theta },{\underline{z%
}}).
\end{multline*}
After a lengthy calculation \cite{BK} we obtain for the case of the lowest
breather the one-breather-$(n-2)$-soliton form factor 
\begin{multline*}
\mathcal{O}_{3\dots n}(\xi ,\underline{\theta }^{\prime
})=\prod_{2<i}F_{sb}(\xi -\theta _{i})\prod_{2<i<j}F(\theta
_{ij})\,\sum_{l=0}^{1}(-1)^{l}\prod_{2<i}\rho (\xi -\theta _{i},l) \\
\times \int_{\mathcal{C}_{\underline{\theta }}}dz_{2}\cdots \int_{\mathcal{C}%
_{\underline{\theta }}}dz_{m}\,\prod_{1<j}\chi (\xi -z_{j},l) \\
\times \prod_{2<i}\prod_{1<j}\phi (\theta _{i}-z_{j})\prod_{1<i<j}\tau
(z_{ij})\,\tilde{p}^{\mathcal{O}}(\xi ,\underline{\theta }^{\prime },z^{(l)},%
{\underline{z}}^{\prime })\,\Psi _{3\dots n}(\underline{\theta }^{\prime },{%
\underline{z}}^{\prime })
\end{multline*}
with ${\underline{z}}^{\prime }=\left( z_{2},\dots ,z_{m}\right) $. The
soliton-breather form factor has been introduced as 
\begin{eqnarray*}
F_{sb}(\theta ) &=&K_{sb}(\theta )\sin \tfrac{1}{2i}\theta
\,f_{sb}^{(0)}(\theta ) \\
f_{sb}^{(0)}(\theta ) &=&\,\exp \int_{0}^{\infty }\frac{dt}{t}\,2\frac{\cosh 
\frac{1}{2}\nu t}{\cosh \frac{1}{2}t}\,\frac{1-\cosh t(1-\theta /(i\pi ))}{%
2\sinh t}\, \\
K_{sb}(\theta ) &=&\frac{-\cos \tfrac{\pi }{4}(1-\nu )/E(\frac{1}{2}(1-\nu ))%
}{\sinh \frac{1}{2}(\theta -\frac{i\pi }{2}(1+\nu ))\sinh \frac{1}{2}(\theta
+\frac{i\pi }{2}(1+\nu ))}
\end{eqnarray*}
The normalization has been chosen such that $F_{sb}(\infty )=1$. The
function $E(\nu )$ was used in \cite{KW,BFKZ} Also we have introduced the
short notations 
\begin{eqnarray*}
\rho (\xi ,l) &=&(-1)^{l}\frac{\sinh \tfrac{1}{2}\left( \xi -\tfrac{i\pi }{2}%
(1+(-1)^{l}\nu )\right) }{\sinh \tfrac{1}{2}\xi } \\
\chi (\xi ,l) &=&(-1)^{l}\frac{\sinh \frac{1}{2}(\xi +\frac{i\pi }{2}%
(1+(-1)^{l}\nu ))}{\sinh \frac{1}{2}(\xi -\frac{i\pi }{2}(1+(-1)^{l}\nu ))}
\end{eqnarray*}
The following identities have been used 
\begin{eqnarray*}
F(\theta _{1}-\theta _{i})F(\theta _{2}-\theta _{i})\tilde{\phi}(\theta
_{i}-z^{(l)}) &=&\,F_{sb}(\xi -\theta _{i})\rho (\xi -\theta _{i},l) \\
\phi (\theta _{1}-z_{j})\phi (\theta _{2}-z_{j})S_{sb}(\xi -z_{j})\tau
(z^{(l)}-z_{j}) &=&\chi (\xi -z_{j},l)
\end{eqnarray*}
for $\theta _{1/2}=\xi \pm \frac{i\pi }{2}(1-\nu ),\,z^{(l)}=\xi -\frac{i\pi 
}{2}(1-(-1)^{l}\nu ).$ The new p-function is obtained from the old one by 
\[
\tilde{p}^{\mathcal{O}}(\xi ,\underline{\theta }^{\prime },z^{(l)},{%
\underline{z}}^{\prime })\,=m\,d(\nu )\,p^{\mathcal{O}}(\xi +\tfrac{1}{2}%
iu^{(1)},\xi -\tfrac{1}{2}iu^{(1)},\underline{\theta }^{\prime },z^{(l)},{%
\underline{z}}^{\prime }\,) 
\]
where the constant $d(\nu )$ is given by 
\[
d(\nu )=\frac{\sqrt{E(\nu )}}{\varkappa \sqrt{\sin \frac{1}{2}\pi \nu }} 
\]

Iterating the procedure above we obtain the $r$-breather-$s$-soliton form
factor with $2r+s=n,$ the breather rapidities $\underline{\xi }=(\xi
_{1},\dots ,\xi _{r})$ and the soliton rapidities$\,\underline{\theta }%
=(\theta _{1},\dots ,\theta _{s})$ 
\begin{eqnarray*}
\mathcal{O}_{1\dots s}(\underline{\xi },\underline{\theta }) &=&\prod_{1\leq
i<j\leq r}F_{bb}(\xi _{ij})\prod_{i=1}^{r}\prod_{j=1}^{s}F_{sb}(\xi
_{i}-\theta _{j})\prod_{1\leq i<j\leq s}F(\theta _{ij})\, \\
&&\times \sum_{l_{1}=0}^{1}\dots \sum_{l_{r}=0}^{1}(-1)^{l_{1}+\dots
+l_{r}}\prod_{1\leq i<j\leq r}\left( 1+(l_{i}-l_{j})\frac{i\sin \pi \nu }{%
\sinh \xi _{ij}}\right) \\
&&\times \prod_{i=1}^{r}\prod_{j=1}^{s}\rho (\xi _{i}-\theta _{j},l)\int
dz_{r+1}\cdots \int dz_{m}\,\prod_{i=1}^{r}\prod_{j=r+1}^{m}\chi (\xi
_{i}-z_{j},l) \\
&&\times \prod_{i=1}^{s}\prod_{j=r+1}^{m}\phi (\theta
_{i}-z_{j})\prod_{r<i<j\leq m}\tau (z_{ij})\,\tilde{p}(\underline{\xi },%
\underline{\theta },\underline{z^{(l)}},{\underline{z}})\,\Psi _{1\dots s}(%
\underline{\theta },{\underline{z}})
\end{eqnarray*}
again with $\,z_{i}^{(l_{i})}=\xi _{i}-\frac{i\pi }{2}(1-(-1)^{l_{i}}\nu
),(i=1,\dots ,r)$. The two-breather form factor has been introduced as 
\begin{eqnarray*}
F_{bb}(\xi ) &=&K_{bb}(\xi )\,\sin \tfrac{1}{2i}\xi \,\,f_{bb}^{(0)}(\xi ) \\
f_{bb}^{(0)}(\theta ) &=&\,\exp \int_{0}^{\infty }\frac{dt}{t}\,2\frac{\cosh
(\frac{1}{2}-\nu )t}{\cosh \frac{1}{2}t}\,\frac{1-\cosh t(1-\theta /(i\pi ))%
}{2\sinh t}\, \\
K_{bb}(\theta ) &=&\frac{-\cos \frac{1}{2}\pi \nu /E(\nu )}{\sinh \frac{1}{2}%
(\theta -i\pi \nu )\sinh \frac{1}{2}(\theta +i\pi \nu )}
\end{eqnarray*}
The normalization has been chosen such that $F_{bb}(\infty )=1$. It has been
used that 
\begin{multline*}
F_{sb}(\xi _{1}-\theta _{3})F_{sb}(\xi _{1}-\theta _{4})\,\rho (\xi
_{1}-\theta _{3},l_{1})\rho (\xi _{1}-\theta _{4},l_{1})\chi (\xi
_{1}-z_{2}^{(l_{2})},l_{1}) \\
=F_{bb}(\xi _{12})\left( 1+(l_{1}-l_{2})\frac{i\sin \pi \nu }{\sinh \xi _{12}%
}\right)
\end{multline*}
for $\theta _{3/4}=\xi _{2}\pm \frac{i\pi }{2}(1-\nu )$. The new p-function
is obtained from the old one by 
\[
\,\tilde{p}(\underline{\xi },\underline{\theta }^{\prime \prime },\underline{%
z^{(l)}},{\underline{z}}^{\prime \prime })\,=\binom{m}{r}r!\,d^{r}(\nu
)\,p\left( \xi _{1}+\tfrac{1}{2}\theta ^{(1)},\xi _{1}-\tfrac{1}{2}\theta
^{(1)},\dots ,\underline{\theta }^{\prime \prime },z_{1}^{(l_{1})},\dots ,{%
\underline{z}}^{\prime \prime }\right) \,. 
\]
In particular for $n=2r=2m$ we get the pure lowest breather form factor 
\begin{equation}
\mathcal{O}(\underline{\xi })=\prod_{i<j}F_{bb}(\xi
_{ij})\,\sum_{l_{1}=0}^{1}\dots \sum_{l_{r}=0}^{1}(-1)^{l_{1}+\dots
+l_{r}}\prod_{1=i<j}^{r}\left( 1+(l_{i}-l_{j})\frac{i\sin \pi \nu }{\sinh
\xi _{ij}}\right) \,\tilde{p}(\underline{\xi },\underline{z^{(l)}})
\label{23}
\end{equation}
and the pure breather p-function 
\[
\tilde{p}(\underline{\xi },\underline{z^{(l)}})\,=r!\,d^{r}(\nu )\,p\left(
\xi _{1}+\tfrac{1}{2}iu^{(1)},\xi _{1}-\tfrac{1}{2}iu^{(1)},\dots
,\,z_{1}^{(l_{1})},\dots \right) . 
\]

\section{The quantum sine-Gordon field equation}

The classical sine-Gordon model is given by the wave equation 
\[
\Box \varphi (t,x)+\frac{\alpha }{\beta }\sin \beta \varphi (t,x)=0. 
\]
and two particle sine-Gordon S-matrix for the scattering of fundamental
bosons (lowest breathers) \cite{KT} 
\[
S(\theta )=\frac{\sinh \theta +i\sin \pi \nu }{\sinh \theta -i\sin \pi \nu } 
\]
where $\theta $ is the rapidity difference defined by $p_{1}p_{2}=m^{2}\cosh
\theta $ and $\nu $ is related to the coupling constant by $\nu =\beta
^{2}/(8\pi -\beta ^{2}).$

From the S-matrix off-shell quantities as arbitrary matrix elements of local
operators are obtained by means of the ``form factor program'' \cite{KW}. In
particular we provide exact expressions for all matrix elements of all
powers of the fundamental bose field $\varphi (t,x)$ and its exponential $%
\mathcal{N}\exp i\gamma \varphi (t,x)$ for arbitrary $\gamma $. Here and in
the following $\mathcal{N}$ denotes normal ordering with respect to the
physical vacuum which means in particular for the vacuum expectation value $%
\langle \,0\,|\,\mathcal{N}\exp i\gamma \varphi (t,x)|\,0\,\rangle =1$. For
the exceptional value $\gamma =\beta $ we find that the operator $\Box ^{-1}%
\mathcal{N}\sin \beta \varphi (t,x)$ is local. Moreover the quantum
sine-Gordon field equation\footnote{%
In the framework of constructive quantum field theory quantum field
equations where considered in \cite{Sch,Fr}.} 
\begin{equation}
\Box \varphi (t,x)+\frac{\alpha }{\beta }\mathcal{N}\sin \beta \varphi
(t,x)=0  \label{e}
\end{equation}
is fulfilled for all matrix elements, if the ``bare'' mass $\sqrt{\alpha }$
is related to the renormalized mass by\footnote{%
Before such a formula was found in \cite{Fa,Za}.} 
\begin{equation}
\alpha =m^{2}\frac{\pi \nu }{\sin \pi \nu }  \label{mass}
\end{equation}
where $m$ is the physical mass of the fundamental boson. The factor $\frac{%
\pi \nu }{\sin \pi \nu }$ modifies the classical equation and has to be
considered as a quantum correction. For the sinh-Gordon model an analogous
quantum field equation has been obtained in \cite{MS}\footnote{%
It should be obtained from (\ref{f}) by the replacement $\beta \rightarrow
ig $. However the relation between the bare and the renormalized mass in 
\cite{MS} differs from the analytic continuation of (\ref{mass}) by a factor
which is $1+O(\beta ^{4})\neq 1$.}. Note that in particular at the `free
fermion point' $\nu \rightarrow 1~(\beta ^{2}\rightarrow 4\pi )$ this factor
diverges, a phenomenon which is to be expected by short distance
investigations \cite{ST}. For fixed bare mass square $\alpha $ and $\nu
\rightarrow 2,3,4,\dots $ the physical mass goes to zero. These values of
the coupling are known to be specific: 1. the Bethe Ansatz vacuum in the
language of the massive Thirring model shows phase transitions \cite{Ko} and
2. the model at these points is related \cite{K3,LeC,Sm1} to Baxters
RSOS-models which correspond to minimal conformal models with central charge 
$c=1-6/(\nu (\nu +1))$ (see also \cite{MS}).

Also we calculate all matrix elements of all higher local currents $%
J_{M}^{\mu }(t,x)$ ($M=\pm 1,\pm 3,\dots $) fulfilling $\partial _{\mu
}J_{M}^{\mu }(t,x)=0$ which is characteristic for integrable models. The
higher charges fulfill the eigenvalue equation 
\begin{equation}
\left( \int dxJ_{M}^{0}(x)-\sum_{i=1}^{n}\left( p_{i}^{+}\right) ^{M}\right)
|\,p_{1},\dots ,p_{n}\rangle ^{in}=0.  \label{J}
\end{equation}
In particular for $M=\pm 1$ the currents yield the energy momentum tensor $%
T^{\mu \nu }=T^{\nu \mu }$ with $\partial _{\mu }T^{\mu \nu }=0$. We find
that its trace fulfills 
\begin{equation}
T_{~\mu }^{\mu }(t,x)=-2\frac{\alpha }{\beta ^{2}}\left( 1-\frac{\beta ^{2}}{%
8\pi }\right) \left( \mathcal{N}\cos \beta \varphi (t,x)-1\right) .
\label{T}
\end{equation}
This formula is consistent with renormalization group arguments \cite{Z,Ca}.
In particular this means that $\beta ^{2}/4\pi $ is the anomalous dimension
of $\cos \beta \varphi $. Again this operator equation is modified by a
quantum correction ($1-\beta ^{2}/8\pi $). Obviously for fixed bare mass
square $\alpha $ and $\beta ^{2}\rightarrow 8\pi $ the model will be
conformal invariant which is related to a Berezinski-Kosterlitz-Thouless
phase transition \cite{KS}. The proofs of the statements (\ref{e}) -- (\ref
{T}) is sketched in the following together with some checks in perturbation
theory. The complete proofs will be published elsewhere \cite{BK}.

A form factor of n fundamental bosons (lowest breathers) is of the form \cite
{KW} 
\[
f_{n}^{\mathcal{O}}(\underline{\theta })=N_{n}^{\mathcal{O}}K_{n}^{\mathcal{O%
}}(\underline{\theta })\prod_{1\leq i<j\leq n}F(\theta _{ij}) 
\]
where $N_{n}^{\mathcal{O}}$ is a normalization constant, $\theta
_{ij}=\theta _{i}-\theta _{j}$ and $F(\theta )$ is the two particle form
factor function. It fulfills Watson's equations 
\[
F(\theta )=F(-\theta )S(\theta )=F(2\pi i-\theta ) 
\]
with the S-matrix given above. Explicitly it is given by the integral
representation \cite{KW} 
\[
F(\theta )=N\exp \int_{0}^{\infty }\frac{dt}{t}\,\frac{\left( \cosh \frac{1}{%
2}t-\cosh (\frac{1}{2}+\nu )t\right) \left( 1-\cosh t(1-\frac{\theta }{i\pi }%
)\right) }{\cosh \frac{1}{2}t\sinh t} 
\]
normalized such that $F(\infty )=1$. The K-function $K_{n}^{\mathcal{O}}(%
\underline{\theta })$ is meromorphic, symmetric and periodic (under $\theta
_{i}\rightarrow \theta _{i}+2\pi i$).

From (\ref{23}) we have 
\begin{equation}
K_{n}^{\mathcal{O}}(\underline{\theta })=\sum_{l_{1}=0}^{1}\dots
\sum_{l_{n}=0}^{1}(-1)^{l_{1}+\dots +l_{n}}\prod_{1\leq i<j\leq n}\left(
1+(l_{i}-l_{j})\frac{i\sin \pi \nu }{\sinh \theta _{ij}}\right) p_{n}^{%
\mathcal{O}}(\underline{\theta },\underline{z})  \label{K}
\end{equation}
where $z_{i}=\theta _{i}-\frac{i\pi }{2}\left( 1+(2l_{i}-1)\nu \right) $.
The dependence on the operator is encoded in the 'p-function' $p_{n}^{%
\mathcal{O}}$. It is separately symmetric with respect to the variables $%
\underline{\theta }$ and $\underline{z}$ and has to fulfill some simple
conditions in order that the form factor function $f_{n}^{\mathcal{O}}$
fulfill some properties \cite{KW,Sm}. These properties follow (see \cite
{BFKZ}) from general LSZ-assumptions and in additions specific features
typical for integrable field theories. In particular the recursion relation
holds 
\begin{equation}
\limfunc{Res}_{\theta _{12}=i\pi }\,f_{n}^{\mathcal{O}}(\theta _{1},\dots
,\theta _{n})=2i\,f_{n-2}^{\mathcal{O}}(\theta _{3},\dots ,\theta
_{n})\left( \mathbf{1}-S(\theta _{2n})\dots S(\theta _{23})\right) .
\label{iii}
\end{equation}
Here we will not provide more details but only give some examples of
operators and their corresponding p-functions:

\begin{enumerate}
\item  The correspondence of exponentials of the field and their p-function%
\footnote{%
For the sinh-Gordon model an analogous representation as (\ref{K}) together
with this p-function was obtained in \cite{BL} by different methods.} is 
\begin{equation}
\mathcal{N}e^{i\gamma \varphi }\leftrightarrow
\prod_{i=1}^{n}e^{(2l_{i}-1)i\pi \nu \gamma /\beta }  \label{q}
\end{equation}
for an arbitrary constant $\gamma $.

\item  Taking derivatives of this formula with respect to $\gamma $ we get
for the field and its powers 
\begin{equation}
\mathcal{N}\varphi ^{N}\leftrightarrow \left(
\sum_{i=1}^{r}(2l_{i}-1)\right) ^{N}.  \label{N}
\end{equation}

\item  Higher currents (for $M=\pm 1,\pm 3,\dots $) correspond to the
p-functions 
\[
J_{M}^{\pm }\leftrightarrow \sum_{i=1}^{n}e^{\pm \theta
_{i}}\sum_{i=1}^{n}e^{Mz_{i}} 
\]
for $n=$ even and zero for $n=$ odd. For $M=\pm 1$ we get the light cone
components of the energy momentum tensor $T^{\rho \sigma }=J_{\sigma }^{\rho
}$ with $\rho ,\sigma =\pm $ (see also \cite{MS}).
\end{enumerate}

In order to prove equations as for example (\ref{e}) and (\ref{T}) we
consider the corresponding p-functions and their K-functions defined by (\ref
{K}). The K-functions are rational functions of the $x_{i}=e^{\theta _{i}}$.
We analyze the poles and the asymptotic behavior and find identities by
using induction and Liouville's theorem. Transforming these identities to
the corresponding form factors one finds the field equation (\ref{e}) and
the trace equation (\ref{T}) up to normalizations.

Normalization constants are obtained in the various cases by the following
observations:

\begin{enumerate}
\item[a)]  The normalization a field annihilating a one-particle state is
given by the vacuum one-particle matrix element, in particular for the
fundamental bose field one has 
\[
\langle \,0\,|\,\varphi (0)\,|\,p\,\rangle =\sqrt{Z^{\varphi }} 
\]
with the finite wave function renormalization constant calculated in \cite
{KW} as 
\[
Z^{\varphi }=(1+\nu )\frac{\frac{\pi }{2}\nu }{\sin \frac{\pi }{2}\nu }\exp
\left( -\frac{1}{\pi }\int_{0}^{\pi \nu }\frac{t}{\sin t}dt\right) 
\]
where $Z^{\varphi }=1+O(\beta ^{4})$. This gives the normalization constant 
\begin{equation}
N_{1}^{(1)}=\sqrt{Z^{\varphi }}/2  \label{N1}
\end{equation}
for the form factors of the fundamental bose field and which are obtained
from the p-function of (\ref{N}) for $N=1$.

\item[b)]  If a local operator is connected to an observable like a charge $%
Q=\int dx\,\mathcal{O}(x)$ we use the relation 
\[
\langle \,p^{\prime }\,|\,Q\,|\,p\,\rangle =q\langle \,p^{\prime
}\,|\,\,p\,\rangle . 
\]
For example for the higher charges we obtain 
\[
N_{2}^{J_{M}}=\frac{i^{M}m^{M+1}}{2\sin \pi \nu \sin \frac{M}{2}\pi \nu
F(i\pi )}\quad \mathrm{with}\quad \frac{1}{F(i\pi )}=Z^{\varphi }\frac{\beta
^{2}}{8\pi \nu }\frac{\sin \pi \nu }{\pi \nu }. 
\]

\item[c)]  We use Weinberg's power counting theorem for bosonic Feynman
graphs \cite{BK}\footnote{%
This type of arguments has been also used in \cite{KW,FMS,KM,MS}.}. For the
exponentials of the boson field $\mathcal{O=N}e^{i\gamma \varphi }$ this
yields in particular the asymptotic behavior 
\[
f_{n}^{\mathcal{O}}(\theta _{1,}\theta _{2,}\dots )=f_{1}^{\mathcal{O}%
}(\theta _{1})\,f_{n-1}^{\mathcal{O}}(\theta _{2,}\dots )+O(e^{-\limfunc{Re}%
\theta _{1}}) 
\]
as $\limfunc{Re}\theta _{1}\rightarrow \infty $ in any order of perturbation
theory. This behavior is also assumed to hold for the exact form factors.
Applying this formula iteratively we obtain from (\ref{K}) with (\ref{q})
the following relation for the normalization constants of the operators $%
\mathcal{N}e^{i\gamma \varphi }$%
\[
N_{n}^{\gamma }=\left( N_{1}^{\gamma }\right) ^{n}\quad (n\geq 1). 
\]

\item[d)]  The recursion relation (\ref{iii}) relates $N_{n+2}$ and $N_{n}.$
For all p-functions mentioned above we obtain 
\[
N_{n+2}=N_{n}\frac{2}{\sin \pi \nu F(i\pi )}\quad (n\geq 1). 
\]
\end{enumerate}

Using c) and d) we calculate the normalization constants for the exponential
of the field $\mathcal{N}e^{i\gamma \varphi }$ and obtain 
\begin{equation}
N_{1}^{\gamma }=\sqrt{Z^{\varphi }}\frac{\beta }{2\pi \nu }.  \label{Nq}
\end{equation}

The normalization constants (\ref{N1}) and (\ref{Nq}) now yield the field
equation (\ref{e}) with the mass relation (\ref{mass}). The statement (\ref
{T}) is proved similarly. The eigen value equation (\ref{J}) is obtained by
taking the scalar products with arbitrary out-states and by using a general
crossing formula \cite{BK}.

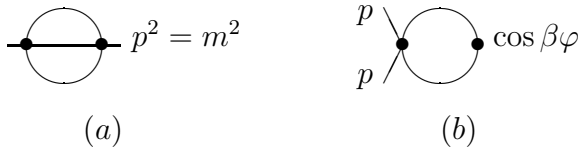
\begin{figure}[tbh]
\[
\begin{array}{l}
\unitlength5mm \begin{picture}(14,3) \put(1,2){\oval(2,2)}
\put(0,2){\makebox(0,0){$\bullet$}} \put(2,2){\makebox(0,0){$\bullet$}}
\put(-.5,2){\line(1,0){3}} \put(2.8,2){$p^2=m^2$} \put(1.5,-.5){$(a)$}
\put(11,2){\oval(2,2)} \put(10,2){\makebox(0,0){$\bullet$}}
\put(12,2){\makebox(0,0){$\bullet$}} \put(12.4,2){$\cos \beta \varphi$}
\put(10,2){\line(-1,2){.5}} \put(10,2){\line(-1,-2){.5}} \put(8.8,1){$p$}
\put(8.8,2.7){$p$} \put(11,-.5){$(b)$} \end{picture}
\end{array}
\]
\caption{Feynman graphs}
\label{f8}
\end{figure}
All the results may be checked in perturbation theory by Feynman graph
expansions. In particular in lowest order the relation between the bare and
the renormalized mass (\ref{mass}) is given by Figure 1 (a). It had already
been calculated in \cite{KW} and yields 
\[
m^{2}=\alpha \left( 1-\frac{1}{6}\left( \frac{\beta ^{2}}{8}\right)
^{2}+O(\beta ^{6})\right) 
\]
which agrees with the exact formula above. Similarly we check the quantum
corrections of the trace of the energy momentum tensor (\ref{T}) by
calculating the Feynman graph of Figure 1 (b) with the result again taken
from \cite{KW} as 
\[
\langle \,p\,|\,\mathcal{N}\cos \beta \varphi -1\,|\,p\,\rangle =-\beta
^{2}\left( 1+\frac{\beta ^{2}}{8\pi }\right) +O(\beta ^{6}). 
\]
This again agrees with the exact formula above since the normalization given
by eq.~(\ref{J}) implies $\langle \,p\,|\,T_{~\mu }^{\mu }|\,p\,\rangle
=2m^{2}$. All other equations have also been checked in perturbation theory 
\cite{BK2}.

\paragraph{Acknowledgements:}

We thank A.A. Belavin, J. Balog, R. Flume, A. Fring, R.H. Poghossian, F.A.
Smirnov, R. Schrader, B. Schroer and Al.B. Zamolodchikov for discussions.
One of authors (M.K.) thanks E. Seiler and P. Weisz for discussions and
hospitality at the Max-Planck Insitut f\"{u}r Physik (M\"{u}nchen), where
parts of this work have been performed. H.B. was supported by DFG,
Sonderforschungsbereich 288 `Differentialgeometrie und Quantenphysik' and
partially by grants INTAS 99-01459 and ISTC A102.

\end{document}